\documentstyle[11pt,newpasp,twoside,epsf]{article}

\def\RCS$#1: #2 ${\expandafter\def\csname RCS#1\endcsname{#2}}
\RCS$Revision: 1.8 $ \RCS$Date: 2000/04/13 23:14:06 $

\begin{document}

\title{Source Reconstruction as an Inverse Problem}
\author{Norman Gray}
\affil{Department of Physics and Astronomy, University of Glasgow,
Glasgow, G12 8QQ, UK}
\author{Iain J Coleman}
\affil{British Antarctic Survey, Madingley Road, Cambridge, CB3 0ET, UK}

\begin{abstract}
Inverse Problem techniques offer powerful tools which deal naturally
with marginal data and asymmetric or strongly smoothing kernels, in
cases where parameter-fitting methods may be used only with some
caution.  Although they are typically subject to some bias, they can
invert data without requiring one to assume a particular model for the
source.  The Backus-Gilbert method in particular concentrates on the
tradeoff between resolution and stability, and allows one to select an
optimal compromise between them.  We use these tools to analyse the
problem of reconstructing features of the source star in a
microlensing event, show that it should be possible to obtain useful
information about the star with reasonably obtainable data, and note
that the quality of the reconstruction is more sensitive to the number
of data points than to the quality of individual ones.

% Revision \RCSRevision, \RCSDate.
\end{abstract}

\section{Introduction}

Where once all the interest in microlensing was in the details of the
lensing population, there is now an increasing interest in lensing events
as probes of the stellar sources.  From this point of view, once the
lens' geometrical details have been worked out, the event can be used
as a `super-telescope', providing otherwise completely unattainable
resolution of the surfaces of distant stellar disks.

Initially, analyses assumed that the microlensing source star could be
taken to be a point source, and the first discussion of `finite source
effects' was in the context of a problem -- Witt \& Mao (1994) asked at
what point the point-source approximation would break down;
{Nemiroff} \& {Wickramasinghe} (1994) put the question more positively, asking what
information could be obtained from the distortions to the light curve
which finite-source effects would cause.

It is not merely intensity information which can be obtained from
events.  Simmons, Newsam, \& Willis (1995) discuss the information which can be
extracted when polarization measurements are made of microlensing
events, and in (Newsam {et~al.} 1998) show how even relatively poor
polarization data can substantially improve fits of source parameters.

Although the basic microlensing effect is achromatic, the fact that
stars have different limb-darkening profiles in different colours
means that a lens differentially amplifying the disk will produce a
chromatic effect.  Other workers
({Valls-Gabaud} 1995; Sasselov 1996; Valls-Gabaud 1998) have discussed how
one might obtain such chromaticity information.  It is even possible
to discuss how one might observe the signatures of stellar spots
(Heyrovsk{\'y} \& Sasselov 1999; Bryce \& Hendry 2000).

The usual way in which source structure is detected is by applying a
model-fitting (equivalently, parameter-fitting) algorithm to the
observed data, to obtain the best-fit parameters of a suitable
limb-darkening model; this is the approach used, for example, by the
MACHO collaboration ({Alcock} {et~al.} 1997) and the PLANET collaboration
({Albrow} {et~al.} 1999) to make the first detections of limb-darkening in
microlensing events.  It is also the approach which underlies the
insightful error analysis by {Gaudi} \& {Gould} (1999).

A parameter-fitting algorithm essentially consists of a mechanism for
systematically moving through parameter space, repeatedly solving the
`forward problem' -- calculating the data to be predicted from a given
limb-darkening profile -- until the predicted data is optimally close
to the data actually observed.  Here we want to suggest that, because
of the fact that the underlying source function is convolved through a
broad and asymmetric amplification kernel, a model-fitting approach is
potentially problematic, and that this recovery problem is more
naturally addressed using the well-established technology of inverse
problems.

We plan to discuss the merits of inverse problem techniques in general,
and the Backus-Gilbert method in particular, and exemplify the
possibilities by inverting simulated microlensing data to recover
limb-darkening and limb-polarization effects.  We will see that we are
able to discuss explicitly and robustly the tradeoff between
resolution and stability which is implicit in any such inversion,
including recoveries obtained by model-fitting.

\section{Background}

\begin{figure}
%\plotfiddle{geom.1}{5cm}{0}{150}{150}{-100}{0}
\plotfiddle{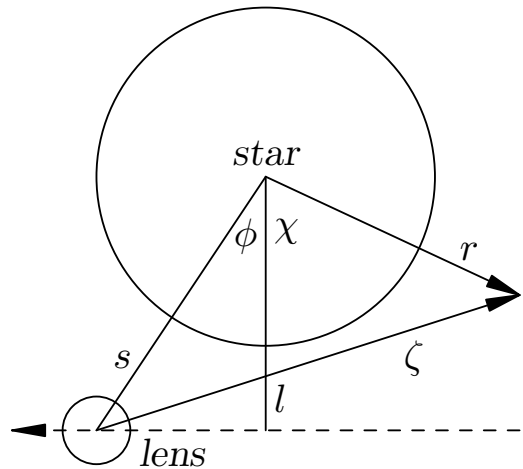}{5cm}{0}{150}{150}{-100}{0}
\caption{\label{f:geom}Geometry of a lensing event.  The projected
path of the lens has impact parameter~$l$, and the path is
parameterised by polar coordinates~$s(t)$ and~$\phi(t)$, relative to
the centre of the source, projected into the lens plane.  Any point in
that plane can be given in polar coordinates~$r$ and~$\chi$, and this
point is a distance~$\zeta$ from the centre of the lens.  All
dimensions are normalised to the Einstein radius in the source plane.
The angles~$\phi$ and~$\chi$ are taken with respect to the line
joining the source to the lens' point of closest approach.}
\end{figure}

The geometry we consider is as shown in Fig.~\ref{f:geom}.
The amplification function is the familiar one.
\begin{equation}
A(\xi) = \frac12\left(\xi+\frac1\xi\right),\quad
\xi = \left(1+\frac4{\zeta^2}\right)^{1/2},\quad
\zeta^2 = r^2 + s^2 - 2rs\cos(\chi-\phi).
\end{equation}

Denote the intensity on the stellar surface~$I(r)$, and the Stokes parameter
by~$Q(r,\chi)=-P(r)\cos2\chi$, where~$P(r)$ is the polarization of the
stellar surface and we are assuming that the surface is
rotationally symmetric.  In the case of a microlensing event, we
cannot directly resolve details of the lensed source, and must therefore
measure integrals over the source surface.  We immediately obtain
\begin{eqnarray}
I(s(t),\phi(t)) &=& \int_0^\infty I(r)\tilde A_I(r;s,\phi)\,{\mathrm{d}} r
	\label{e:IIP}\\
Q(s(t),\phi(t)) &=& \int_0^\infty P(r)\tilde A_Q(r;s,\phi)\,{\mathrm{d}} r
	\label{e:QIP}
\end{eqnarray}
where the amplification kernels are
\begin{eqnarray}
\tilde A_I &=& r\int_0^{2\pi} A(r,\chi;s,\phi)\,{\mathrm{d}} \chi 
	\label{e:Idef}\\
\tilde A_Q &=& -r\int_0^{2\pi} \cos2\chi A(r,\chi;s,\phi)\,{\mathrm{d}}\chi.
	\label{e:Qdef}
\end{eqnarray}
Note that the kernel~$\tilde A_I$ is a factor~$2\pi r$ times the
angular average of the amplification function, and the
functions~$I(s,\phi)$ and~$Q(s,\phi)$ have the dimensions of flux
rather than intensity.

Analytic expressions for these angle-averaged amplification functions
have been obtained by numerous people.  Schneider \& Wagoner (1987) and
{Gaudi} \& {Gould} (1999) perform the average for approximate forms of the
amplification function, {Heyrovsk{\'y}} \& {Loeb} (1997) deal with an elliptical
source for a slightly restricted class of source functions.
Gray (2000) produced integrals for the exact amplification
function, for axisymmetric sources, obtaining
\[
\tilde A_I(r;s,\phi) = r(2\pi + I_1 - I_2), \qquad
\tilde A_Q(r;s,\phi) = -r \cos 2\phi (Q_1 - Q_2)
\]
where~$I_1$, $I_2$, $Q_1$ and $Q_2$ are elliptic integrals whose
arguments have an algebraic dependence on~$r$ and~$s$.  Although it is
helpful, it is not in fact necessary to have analytic forms for the
angle-averaged amplification functions, and the following analysis
would be just as successful if these could be obtained only
numerically.

Equations~(\ref{e:IIP}) and~(\ref{e:QIP}) are one-dimensional integral
equations -- the classic form of an inverse problem.  In this context,
the function~$I(r)$ is termed the underlying or source function, and
the function~$\tilde A_I(r;s)$ the kernel.  The data is the set of
values~$I(s(t_i),\phi(t_i))$ for some set of times~$t_i$.

The use of inverse problem techniques is relatively rare in this
branch of astrophysics.  Mineshige \& Yonehara (1999) used the technique to map
the einstein cross accretion disk using a hypothetical caustic
crossing, and Wambsganss (2000) points out a number of other uses
in the field of cosmological microlensing.  Coleman, Gray, \& Simmons (1998) use a
similar technique with eclipsing binaries.  The method described in
this paper is discussed at greater length in Gray \& Coleman (2000).

\section{Inverse problem techniques vs. parameter fitting}

Parameter fitting is the most appropriate technique in the case where
(a) there is no doubt about the most appropriate model to use, so that
the aim is simply to recover model parameters, and (b) when the
problem kernel is not {\em ill-conditioned}.  When the model itself is
open to dispute, or the observational situation means that the kernel
is ill-conditioned, then any parameter fit must be done extremely
cautiously if it is not to be deceptive.

Ill-conditioning can be characterised in several ways.  Fundamentally,
an ill-conditioned kernel maps a large volume of parameter space to a
small volume of data space; a strongly smoothing (nearly flat) kernel
would be an example of this.  This implies that the inversion is
highly unstable, so that a tiny, noise-induced, change in the data
($I(t)$ in Eqn.~(\ref{e:IIP})) could, after a na{\"\i}ve inversion, be
taken to indicate a radically different recovered function~$I(r)$.

Inverse problem techniques -- also known as `non-parametric' or
`model-free' techniques -- dispense with a parameterised model, and
instead approach the problem from the question `how much information
does this kernel permit to be recovered from this data?' (see, for
example, Craig \& Brown (1986) for a general introduction).  They are most
natural in the case of marginal data, or an asymmetric or broad
kernel.  These techniques are typically associated with Bayesian
approaches to data analysis, and deal with the instability of the
inversion by adding prior information, such as the supposition that
the underlying function be smooth (in the case of inversion by
regularisation) or otherwise featureless (in the case of
maximum-entropy inversion).  This explicit addition of prior
information inevitably makes inverse problem techniques suffer from
bias; in a parameter-fitting algorithm, the model acts as implicit
prior information, so that route is not free of bias either.

\subsection{Backus-Gilbert}

The particular technique we have used is the Backus-Gilbert method.
This works by allowing us to explicitly trade off recovery resolution
against stability.  We very briefly outline the method here; there is a
fuller account in, for example, Parker (1977), and an astrophysical
example in Loredo \& Epstein (1989).

Given a kernel~$K(r;s_i)$, underlying function~$u(r)$, data $F(s_i)$,
and noise~$n_i$, the general 1-d inverse problem can be written as
\[
F(s_i) = \int u(r) K(r, s_i) \,{\mathrm{d}} r + n_i.
\]
(compare Eqns.~(\ref{e:IIP}) and~(\ref{e:QIP})).
We suppose that we can find `response kernels' $q_i(r)$ which permit
us to form an estimate~$\hat u$ of the underlying function as a
weighted average of the data:
\begin{equation}
\hat u(r) = \sum_i q_i(r) F(s_i).
\label{e:uqF}
\end{equation}
This is a random variable, but we can
relate its mean to the underlying function through an
`averaging kernel' $\Delta$:
\begin{equation}
\langle\hat u(r)\rangle = \int \Delta(r,r') u(r') \,{\mathrm{d}} r'.
\label{e:Delta}
\end{equation}
Ideally, this kernel would be the Dirac delta function, and the
estimator~$\hat u$ would perfectly track the underlying function.
Since the underlying function is (of course) unknown, we cannot use
Eqn.~(\ref{e:Delta}) directly, but this definition of~$\Delta(r,r')$
allows us to define 
$\mathop{\mathrm{Width}}[\Delta]$ and $\mathop{\mathrm{Var}}[\hat
u(r)]$, which depend only on the kernel~$K(r;s_i)$ and the
noise~$n_i$.  This means that we can explicitly trade off improved
recovery resolution (narrower 
$\mathop{\mathrm{Width}}[\Delta]$) against improved stability (smaller
$\mathop{\mathrm{Var}}[\hat u(r)]$), with the relative weighting
parametrised by smoothing parameter~$\lambda$.  For each value
of~$\lambda$ we can analytically obtain a set of coefficients~$q_i(r)$
for Eqn.~(\ref{e:uqF}).

Note that this technique is an analysis of the {\em kernel}, rather
than a particular data set, which means (a) the understanding we gain
of how much information is available from a data set is portable both to
other inverse problem techniques, and also to parameter-fitting
approaches, and (b) the analysis can be done, and an optimal~$\lambda$
selected, {\em prior to any data being collected}, given a set of
demands on the required resolution and stability.  A feature of the
Backus-Gilbert method is that the response kernels~$q_i(r)$ has a
dependence on the radial parameter~$r$, so that the analysis needs to
be redone for each~$\hat u(r)$ we wish to recover (at a cost of
potentially large matrix inversions each time).  This means that
the Backus-Gilbert method (at least in its simplest form) is not an
obvious choice for a data reduction pipeline, but it has the
compensation that the optimal tradeoff can be chosen differently for
different values of~$r$.

\section{Recovery of limb-darkening and limb-polarization profiles}

\begin{figure}
%\plotfiddle{lambda_plot.eps}{7cm}{0}{65}{65}{-180}{-15}
\plotfiddle{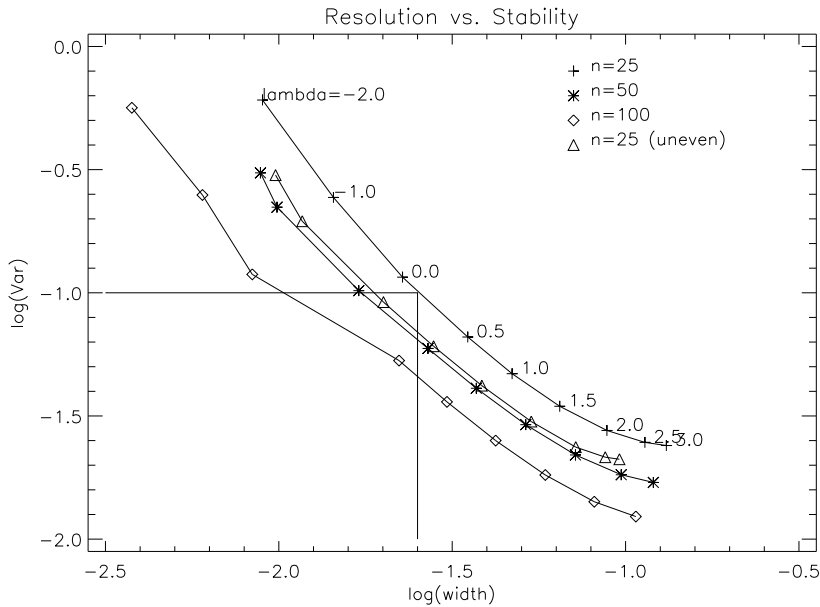}{7cm}{0}{65}{65}{-180}{-15}
\caption{\label{f:lambda}Resolution versus stability for various
numbers of data points, and choices of smoothing parameter~$\lambda$.
The box in the lower left hand
corner shows the region of `acceptable' recoveries.}
\end{figure}

In Figure~\ref{f:lambda}, we show the tradeoff curves for the
polarization kernel, Eqn.~(\ref{e:QIP}), for a variety of numbers of
data points, and choices of smoothing parameter~$\lambda$ (there is a
similar curve for limb-darkening, Eqn.~(\ref{e:IIP})).  Recovery
quality can be improved either by increasing the number of data
points, or by adjusting the times at which data is taken.  In the
figure, the line marked `$n=25$ (uneven)' shows the tradeoff curve for
data taken unevenly, with observations clustered at points the
analysis shows to be particularly sensitive; it is clear that the
quality of the recovery is sensitive to both the number of data points
and the manner in which they are obtained.

The bottom left of the plot corresponds to high resolution and high
stability, the top left to recoveries with high resolution and low
stability (in the limit, picking a single data point and scaling it),
and the bottom right to low resolution and high stability (in the
limit, simply averaging all the data).

The question of what counts as `adequate' resolution and stability can
only be decided in the context of a particular stellar model (though
this does not make the Backus-Gilbert analysis, or the trade-off
diagram in Figure~\ref{f:lambda}, model-dependent).  For a pure
electron atmosphere, Chandrasekhar calculated a limb-polarization of
11\%, suggesting that the variance of the limb-polarization recovery
may be of order 10\% at most.  Estimating the width of a typical
limb-polarization profile suggests a similar upper bound for the
resolution in Figure~\ref{f:lambda}.  The figure therefore indicates
what value of the tradeoff parameter~$\lambda$ we should pick, and
therefore what is the best resolution and variance we can expect to
achieve with a given number of data points.

\begin{figure}
%\plotfiddle{iain-recovery.eps}{7cm}{0}{65}{65}{-180}{-20}
\plotfiddle{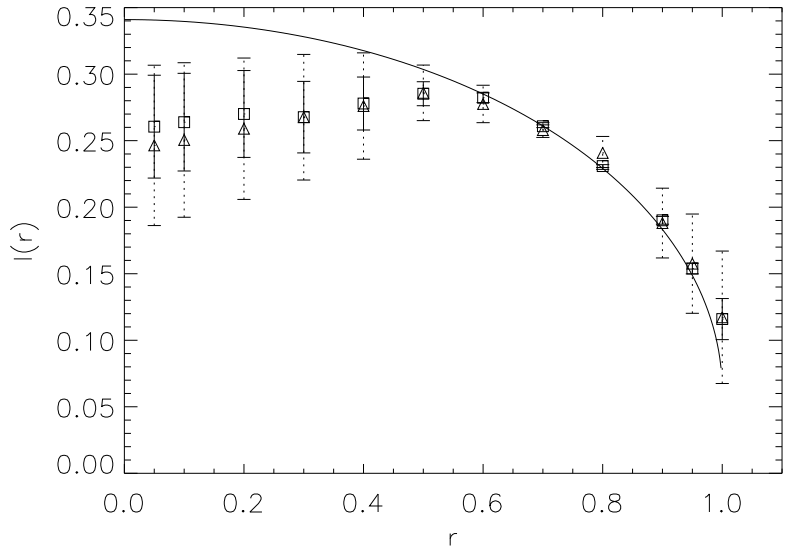}{7cm}{0}{65}{65}{-180}{-20}
\caption{\label{f:iain}Recoveries of a limb-darkening profile
(solid line) with 100 data points, and impact parameter, einstein
radius and stellar radius all equal.  The points plotted with boxes
and solid error bars are recovered from simulated data with
$\sigma=0.002$, and the points with triangles and dotted error bars
from data with $\sigma=0.1$.  After \protect Coleman (1998).}
\end{figure}

In Figure~\ref{f:iain}, we show two recoveries of a limb-darkening
profile from synthetic data with different noise standard-deviation.
Note (a) that the optimal recovered variance varies across the disk;
(b) that the recovery is biased -- towards a flat profile in this case
-- with the bias being greatest at the centre of the disk and at the
limb; (c) that both the bias and the variance are least around 70\% of
the projected radius, which is particularly fortunate since it is in
this region that limb-darkening profiles tend to be most sensitive to
parameters such as~$g$, and where the profiles of giants differ most
from those of normal stars (Hauschildt 2000); and (d) that the
quality of the recovery is surprisingly insensitive to the quality of
the data.

\section{Discussion}

One of the aims of this paper is to emphasise the seriousness of the
ill-conditioning of the source reconstruction problem, and hence the
desirability of using an analytical technique which starts by
examining that ill-conditioning, goes on to discuss what information
is nonetheless recoverable, and only then produces the numerical
information which is the point of the exercise.  Of course, the same
questions can be asked using a parameter-fit approach, but less
naturally, since such approaches assume, in a sense, that the
information is recoverable, with the result that problems can only be
uncovered \textit{post hoc}, by an intelligent examination of
goodness-of-fit measures, or by tracing the propagation of errors
through a calculation.

The second aim is to use this inverse problem approach to analyse the
kernel which turns the underlying limb-darkening and limb-polarization
functions into microlensing intensity and polarization data.  It
turns out that the information is indeed recoverable with adequate
uncertainties but, depending on the quality of the data available, one
many have to make significant compromises over the resolution one is
prepared to accept.  It is also possible to use this approach to
analyse the effect of modifications in the way the data is collected,
and discover that this can indeed significantly improve the data's
effective quality.

\let\JApJ\apj
\let\JMNRAS\mnras
\let\JPASJ\pasj
\let\JAAS\aaps
%\bibliographystyle{apj}
%% \bibliography

\end{document}